\documentclass[twocolumn,showpacs,preprintnumbers,amsmath,amssymb]{revtex4} 
\usepackage{graphicx}
\usepackage{times}

\newcommand{\ba}{\begin{eqnarray}}
\newcommand{\ea}{\end{eqnarray}}
\newcommand{\rt}{$\langle r \rangle_{(2)}$}
\newcommand{\rp}{$r_{rms}$}
\begin{document}
\normalsize
\twocolumngrid
\noindent
{\bf \large Comment on "Constraints on proton structure from precision atomic 
physics measurements"} \\[1mm]

In a recent letter [1] Brodsky {\em et al.} used experimental and
theoretical information on the hyperfine structure (hfs) of hydrogen and muonium
to derive the proton structure correction. In our previous work [2] on this
topic we used only information on the hydrogen hfs to calculate a corresponding
correction, which was consistent with that of [1]. The structure correction can
be decomposed into a dominant static part (the Zemach term, proportional
to the Zemach moment, \rt) and a dynamic part (the polarization correction). In
[2] we directly determined the Zemach term from the (experimental)
electron-scattering form factors that define it. From this result we were able
to {\em infer} the polarization correction. Reference [1] used a value of the
polarization correction calculated in [3] to infer the Zemach term. The
two results for \rt~ are in severe disagreement and we strongly disagree with 
the conclusions of [1].

\begin{figure}[thb]
\begin{center}
\includegraphics[scale=0.46,clip]{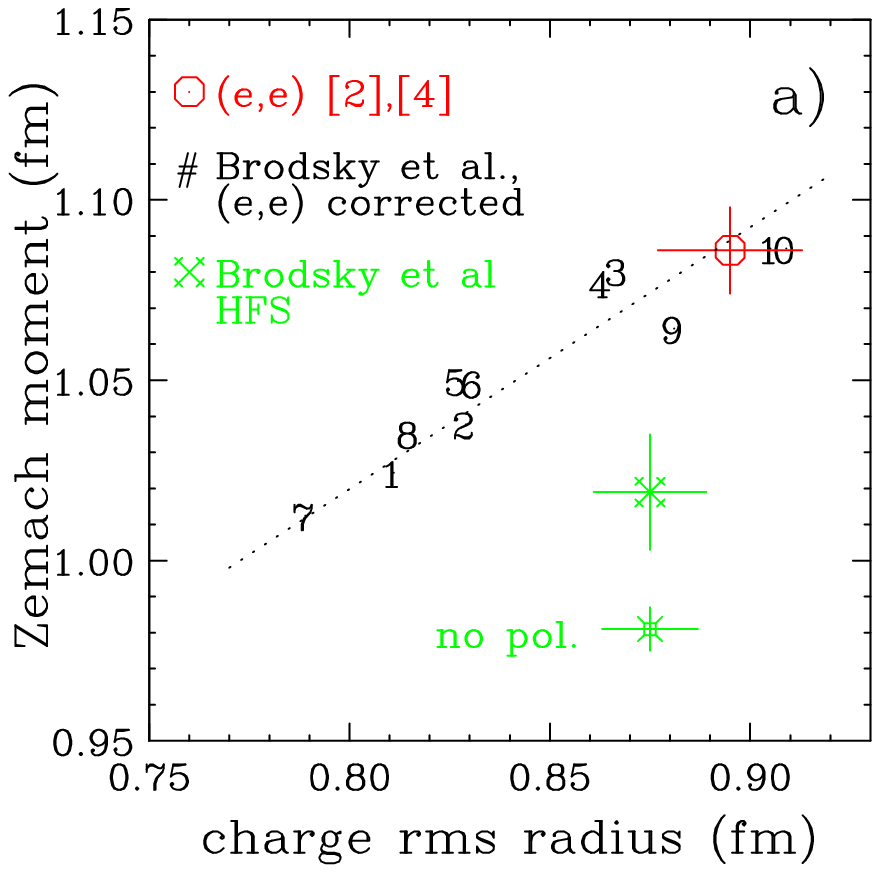}
\includegraphics[scale=0.46,clip]{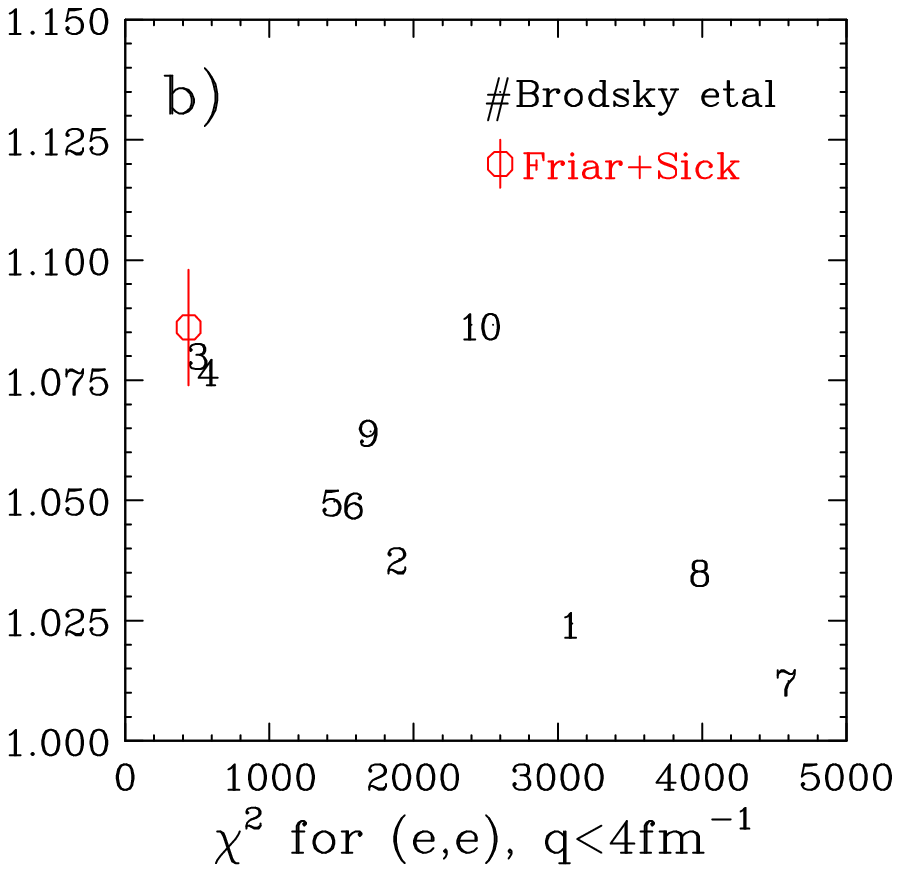} 
\end{center}
\vspace*{-6mm}
\begin{center} {\caption[]{(a) Correlation between \rp~ and \rt ; the
parametrizations of [1] are labeled by \#, and are fit by the dotted line.
(b) $\chi^2$ of the parametrizations
employed in [1].}} 
\end{center} 
\end{figure}
\vspace*{-8mm}
Although the resulting \rt~ of Ref.~[1] disagrees with the precise value from
electron scattering [2], Brodsky {\em et al.} claim that combinations of model
form factors $G_e^p (q)$ and $G_m^p (q)$ exist that are compatible with their
\rt~. This conclusion is not justified. Values of the proton charge $rms$-radius
\rp~ and \rt~ for 10 different combinations of $G_e^p$ and $G_m^p$ are listed in
Table 1 of [1]. All values of \rt~ and several values of \rp~ are numerically
incorrect; the correct values are displayed in Fig.~(1a). This figure shows that
no combination of proton form factors is compatible with the upper green symbol,
which depicts the \rt~ of Ref.~[1] and the \rp~ derived from the hydrogen Lamb
shift [4] (which agrees with electron scattering [5]). 
The figure also clearly shows that there is a tight correlation
between \rp~ and \rt.  The \rt~ of Brodsky {\em et al.} is incompatible with all
other information if any modern value of \rp~ is correct. Our result in red is
compatible.

Table 1 of [1] and the spread of values in Fig.~(1a) might suggest that 
electron scattering does not accurately determine \rt , notwithstanding the
precise result of 1.086$\pm$0.012 fm in [2]. Figure~(1b) shows that this
conclusion is not  justified. For the range of momentum transfers relevant
for the determination of \rt ($q < 4 fm^{-1}$), most of the
$G_e^p,G_m^p$-combinations of [1] give a disastrous $\chi^2$ when compared to
the {\em world} electron scattering data; they simply do not fit that data. The
only exceptions are fits 3 and 4, which differ from the one of Refs.~[2,5] 
mainly by the omission of Coulomb distortion, which explains the slightly low
value of \rp~.

Figure~(1a) also points out the sensitivity of Ref.~[1]'s determination of \rt 
to the  polarization correction (omitting it leads to the lower green symbol). This correction has been calculated by Faustov {\em et al.} [3].
It depends on the proton spin structure functions $g_1(q,x)$ and $g_2(q,x)$ at
low $q$ and low energy loss $\nu$ (large $x$), where these structure functions
are poorly known. Modeling with the constituent quark model (as in [3]) might
introduce errors. The uncertainty  attributed in Ref.~[3] to the polarization
correction  was a
subjective theoretical estimate (unlike the uncertainty in [2]), and we believe
that it has been substantially underestimated.

We conclude that the technique of Brodsky {\em et al.} to use hydrogen and
muonium hfs (together with estimates of the polarization correction) to
calculate \rt will remain inaccurate until the polarization correction can be
calculated using more and better inelastic electron scattering data. Until that
time the most reliable value for \rt~ will come from elastic electron
scattering.  \\[5mm]
J.L. Friar \\
\hspace*{4mm} Los Alamos National Laboratory \\
\hspace*{4mm} Los Alamos, NM 87545 \\[2mm]
I. Sick \\
\hspace*{4mm} University of Basel \\
\hspace*{4mm} CH4056 Basel, Switzerland \\[5mm]
%
DOI \\
PACS numbers: 14.20Dh, 13.40.Gp, 31.30.-i \\[4mm]
$[1]$ S.J. Brodsky {\em et al.}, Phys. Rev. Lett. 94(2005)033002 \\
$[2]$ J.L. Friar and I. Sick, Phys. Lett. B 579(2004)285 \\
$[3]$ R.N. Faustov {\em et al.}, Eur. Phys. J. C24(2002)281 \\
$[4]$ P.J. Mohr {\em et al.}, Rev. Mod. Phys. 77(2005)1\\
$[5]$ I. Sick, Phys. Lett. B576(2003)62 \\

\end{document}